\documentclass[article]{revtex4}
\usepackage{amsmath,amssymb,amsfonts}
\usepackage{graphics,graphicx,psfrag}

\newcommand{\be}{\begin{equation}}
\newcommand{\ee}{\end{equation}}

\begin{document}


\title{Colloidal ionic complexes on periodic substrates: ground state
  configurations and pattern switching}

\author{Samir El Shawish} 
\affiliation{Department of Theoretical Physics, Jo\v zef Stefan Institute,
  Jamova 39, 1000 Ljubljana, Slovenia}

\author{Jure Dobnikar} \affiliation{ Department of Chemistry, University of Cambridge, 
Lensfield Road, CB2 1EW, Cambridge, UK \\ Department of Theoretical Physics, Jo\v
  zef Stefan Institute, Jamova 39, 1000 Ljubljana, Slovenia}

\author{Emmanuel Trizac} \affiliation{Universit\'e Paris-Sud, Laboratoire de
  Physique Th\'eorique et Mod\`eles Statistiques (CNRS UMR 8626), 91405 Orsay
  Cedex, France}

\begin{abstract}

We theoretically and numerically studied ordering of ``colloidal
ionic clusters'' on periodic substrate potentials as those
generated by optical trapping. Each cluster
consists of three charged spherical colloids: two negatively and one
positively charged. The substrate is a square or rectangular array of
traps, each confining one such cluster. By varying the lattice
constant from large to small, the observed clusters are first rod-like
and form ferro- and antiferro-like phases, then they bend into a
banana-like shape and finally condense into a percolated
structure. Remarkably, in a broad parameter range between
single-cluster and percolated structures, we have found stable
super-complexes composed of six colloids forming grape-like or
rocket-like structures. We investigated the possibility of macroscopic
pattern switching by applying external electrical fields.

\end{abstract}

\date{\today}
\maketitle

\section{Introduction}

It is generally considered that the study of colloidal systems started
with the work of Thomas Graham, around 1860, although substances such
as the ``purple of Cassius'', which turns out to be colloidal gold,
were used by glass-makers and porcelain manufacturers as early as the
XVIIth century \cite{Hedges}. For a long period, those systems were
considered as ``unmanageable'' (see e.g. the preface of
Ref. \cite{Hedges}), due in particular to the difficulty to
characterize them precisely (size, shape, charge etc) and also to the
lack of theoretical tools to understand effective interactions.  From
this perspective, a decisive step forward occurred in the 1940s with
the work of Derjaguin, Landau, Verwey and Overbeek \cite{DLVO}, which
helped turn the initial empiricism into a more solid body of
knowledge. More recently, progress in experimental manipulation
techniques promoted colloids as interesting model systems to
investigate a large gamut of fundamental physical phenomena: phase
ordering, nucleation, effects of confinement and reduced
dimensionality, interfacial phenomena, and various non-equilibrium
problems including glass formation, to name but a few. Indeed,
compared to atomic systems, colloids offer four advantages~: first,
they are characterized by small elastic constants and therefore they
are easy to manipulate (``soft matter''); second, they are more easily
visualized since their typical length scale is within the range of
visible light; third, the corresponding time scale is no longer in the
pico-second range, but of order 1s; fourth, their effective
interaction potential may be easily tuned to some extent by changing
some control parameter such as the concentration of a given solute
\cite{Frenkel}. Such features stem, directly or indirectly, from the
large size (on an atomic scale) of the mesoscopic constituents.
Hence, colloids can be considered as ``big atoms'' \cite{Poon}, but
with controllable interactions, and it is noteworthy that they may
furthermore exhibit new structures, that do not seem to have any
atomic analogue \cite{Leunissen}.

Of particular interest here is a class of charged colloidal spheres in
the presence of optical trap arrays, where a wide variety of
crystalline states has been reported
\cite{Bech2002,RO,Agra,Sarlah,ElShawish,Bunch1,ElShawish2}.  On a
two-dimensional modulated substrate, multiple charged colloids can be
confined in sufficiently strong traps, and thereby form $n$-mers which
exhibit an orientational degree of freedom. It has been established
from experimental, numerical and theoretical approaches
\cite{Bech2002,RO,Agra,Sarlah,ElShawish,Bunch1,ElShawish2} that such
systems may show remarkably rich orientational ordering, and lead to
the formation of so called colloidal molecular crystals.

Previous approaches considered situations where the colloids forming
trapped $n$-mers are like-charged objects. On the other hand, the
`molecules' under study here are made up of oppositely charged
colloids, with equal charges in absolute value. The focus will be on a
square lattice of traps, including also distorted geometries where all
distances along one of the principal axis of the square is scaled by a
given factor $\alpha$, thereby producing a rectangular lattice.  We
will restrict our attention to the ground state, leaving thermal
effects for further studies \cite{rque}. Under these circumstances, it
has been shown that in the dipolar case (with trapped clusters each
formed of two oppositely charged colloids) the resulting orientational
behaviour is somehow trivial: the observed phase is made up of
ABAB-type stripes, where the dipoles in stripe A are aligned along the
stripe (up), while those in stripe B have opposite orientation
\cite{ElShawish2}.  Furthermore, the stripe structure cannot be tuned
by modifying the lattice constant or salt content in the solution.
Here we consequently consider clusters with three colloids, two
negatively and one positively charged, which are  more amenable to
an external control, and display a wealth of ordering patterns, which
is the center of our interest. The paper is organized as
follows. The model and the methods are defined in section
\ref{sec:model}, where the differences in modelling with
previous works are outlined. Behaviour on square and rectangular
lattices with isotropic parabolic confinement potential is studied in
section \ref{sec:square} and in section \ref{sec:realtrap}, we discuss
the relevance of these results in the case of a more realistic cosine
trap potential. All these results pertain to equilibrium
configurations, while orientational pattern modifications induced by
an external electric field is finally addressed in section
\ref{sec:electric}. In the conclusion, we summarize our main findings
and discuss perspectives and open problems left.

\section{Model and methods}
\label{sec:model}

\begin{figure}[ht]
\includegraphics[width=0.3\textwidth]{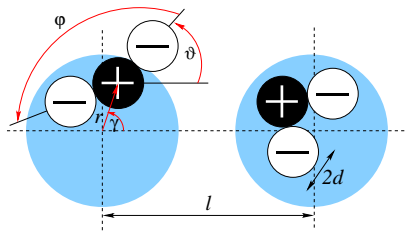}
\caption{(Color online) Sketch of two clusters -- each made up of
  three colloids --
 in two different potential traps, with
  definition of the relevant
 quantities. The center of a blue area
  denotes the confining
 potential minimum and $l$ is therefore the
  intertrap distance. In
 all the paper, the same color code will be
  used: black (resp. white)
 dots for positively (resp. negatively)
  charged colloids.}
\label{S}
\end{figure}

We consider a mixture of negatively and positively spherical and homogeneously
charged colloids, with stoichiometry 2:1 (twice more negative than positive
macroions, which is equivalent to the situation where all charges are reversed). 
Those mesoscopic objects bear a total charge that is neutralized
by an ensemble of micro-ions. We assume that counter-ions and electrolyte
micro-ions do not differ. The colloids are confined in a two dimensional
plane by the action of laser beams. This plane is furthermore corrugated by
additional preferential confinement: the colloids, subject to gradient forces
and light pressure \cite{Bech2002,BB04} tend to gather in the regions of
highest laser intensity. It is then experimentally possible to create a 2D
periodic substrate of traps, with variable geometry.  It should be emphasized
that the micro-ions are not sensitive to confinement, so that our two
dimensional colloidal system is immersed in a standard electrolyte, that
mediates between the charged colloids a screened Coulomb interaction of the
form \cite{DLVO,Hansen,Levin}
\be
   V_C=K\sum_{j\ne i} s_{ij} \frac{e^{-\kappa
   r_{ij}}}{r_{ij}/l}.\label{eq:dlvo}
\ee
The energy scale $K$, which contains a geometric prefactor together with the
squared effective charge of the colloids \cite{ET2002}, is not relevant for
our purposes, since we are interested in ground state configurations. In
Eq. (\ref{eq:dlvo}), the sum runs over all colloidal pairs with interaction sign
$s_{ij}=\pm 1$ distinguishing like- from oppositely charged colloids, and
inter-particle distance $r_{ij}$ between macroions $i$ and $j$; with $l$ we
denote the intertrap distance, see Fig. \ref{S}. 
We therefore implicitly assumed that oppositely
charged colloids at contact also interact through a screened Coulomb form,
whereas such a functional dependence is better suited to describe the
far-field behaviour \cite{Belloni}. We note however that as long as we are
comparing phases with an equal number of doublet +/- at contact, such
configurations contribute a constant to the total energy, so that considering
more exact forms of interaction potential at short scale would not alter our
conclusions. In Eq. (\ref{eq:dlvo}), $1/\kappa$ denotes the Debye length, that
can be tuned changing the electrolyte density. Strictly speaking,
Eq. (\ref{eq:dlvo}) is not an energy but a free energy, since it includes an
entropic contribution from the micro-ions \cite{rque2}.

In addition to inter-particle forces (two-body, and pairwise
additive), the colloids feel an external (one-body) potential due to
laser modulation, and it is the subtle interplay between both kinds of
forces that selects the type of ionic complexes that is formed in each
trap, and their relative arrangement from trap to trap. In
\cite{Agra,Sarlah}, it was assumed that the colloidal complexes in a
given trap (dimers, trimers etc) form a rigid object, with a single
orientational degree of freedom, which effectively allows to omit the
laser-induced potential in the analysis. The situation with oppositely
charged colloids however differs: take an isolated triplet -/+/-, in
the aligned geometry which minimizes its self energy. Such an object
invariably bends under strong confinement thereby forming a
`banana-like' object. The shape of the banana depends on the
parameters (salinity, inter-trap distance etc), and should be
determined self-consistently. We therefore do not lump substrate
potential effects into a unique quantity as in \cite{Agra,Sarlah}, but
explicitly consider each colloidal degree of freedom in the
analysis. This rules out the discrete angle approach invoked in
\cite{Sarlah}, where only a restricted set of possible orientations
were considered.  We also note that already at the level of
like-charged colloids, our recent work \cite{ElShawish} evidenced some
subtle shortcomings of the rigid $n$-mer approach. 

The system obviously bears some resemblance to the molecules on the
atomic scale. However, in our model the binding is the result of the
attractive Yukawa interactions and the confinement potentials and is
therefore relatively weak compared to the strong covalent bonds
typical in real molecules.  We have consequently chosen to call the
$n$-meric colloidal structures ``clusters'' or ``complexes'' rather
than ``molecules'' as has been done elsewhere (see
\cite{Agra,Sarlah,ElShawish,RO}). In spirit, our study then belongs
to a more general line of research pertaining to colloidal
clusters \cite{clusters}. The previous references though generally
deal with large assemblies, whereas our clusters are made up of
a small number of colloids.

In all our considerations, the ratio of the total number of colloids to
the total number of traps is exactly three, therefore under strong
enough confinement, there are exactly 3 colloids per trap and the
ground state is paved with trimers ($- + -$) of varying shapes and
orientations. These mesoscopic objects bear a net charge, which is
neutralized by micro-ions in solution. It has been realized
\cite{Agra,ElShawish} that it is not necessary to invoke local
anisotropies of the confinement potential to account for the observed
ground states: there is an obvious source of anisotropy in the lattice
geometry itself; however, in the vicinity of a trap minimum, the
confined particles feel an isotropic (2D) potential.  In this case,
the screened Coulomb potential alone is able to select preferred
orientations for the clusters (whereas as alluded to above, the
effective spin model considered in \cite{Sarlah} restricted possible
orientations to the principal directions of the light lattice). Here
we wish to avoid introducing an orientational bias through the trap
potential, and therefore initially consider the parabolic
confinement. The total dimensionless energy of colloid $i$ reads
\be
   e_i=A(\delta_i/l)^2 + \sum_{j \neq i} s_{ij}\frac{e^{-\kappa
   r_{ij}}}{r_{ij}/l}, \label{eq:energy}
\ee
where $\delta_i$ denotes the distance between colloid $i$ center of mass and
the center of the trap to which it belongs, and $A$ measures the relative
importance of parabolic confinement over Coulombic forces. Each triplet of
colloids belongs to a `native' trap, so the colloids are not allowed to hop
from one trap to another. Such a limitation precludes the sliding states
studied in \cite{Olson1} under the influence of an external electric field. We
note that the experimental situation is somehow intermediate between the point
of view adopted here -- {\it a priori} mostly relevant for strong confinement
conditions --, and that used in \cite{Sarlah}, see the discussion in
\cite{ElShawish}. For completeness, we have also performed simulations with a
`realistic' potential in the spirit of \cite{RO} to assess the validity of
the isotropic approach.

At constant temperature, three parameters govern the behaviour of the
system: the relative magnitude of the confinement with respect to the
Yukawa interaction $A$, the ratio of the colloid radius to the
Debye screening length
 $\kappa d$, and the ratio of the lattice spacing to Debye length 
 $\kappa l$. In
the rectangular case, the lattice aspect ratio $\alpha$ should be
included. A trimer in two dimensions has six degrees of freedom, but
only four of them are independent assuming the colloids are in
contact. The arrangement of colloids in the vicinity of a trap center
can therefore be described by four quantities (see Fig. \ref{S}): the
position of the cluster's center of mass relative to the trap minimum
is described by $r$ and $\gamma$, the relative orientation of its axis
with respect to a fixed direction (say a line joining
the centers of two neighboring traps) by the angle $\vartheta$ and the amount of
bending by $\phi$ ($\phi=\pi$ for rod-like and $\phi < \pi$ for bent
structures).

Ground state configurations were obtained by energy minimization (EM)
technique and tested by standard Monte Carlo (MC) annealing
simulations. In order to reduce the number of minimizing parameters in
EM, an assumption on the ground state structure was preset. For a
given rectangular lattice we decided to partition the lattice to 2
(A-B) or 4 (A-B-C-D) sub-lattices, which in general form a checkerboard
or a stripe (chain) pattern. Since each colloidal cluster is
parametrized by four quantities (see Fig. \ref{S}), a $p$-partite
($p=2,4$) assumption in EM would correspond to $4p$ minimizing
parameters. We used standard numerical routines (simplex and
quasi-Newton algorithm \cite{nr}) for finding the (local) minimum; by
repeating the EM procedure several ($\sim 10^4$) times over $4p$
different initial estimates, we eventually converged to a global
minimum. However, in extreme parameter regimes (i.e., $l/d\gg 1$ or
close to phase transitions), the convergence became slower due to many
almost degenerate local minima and the ground state was identified
among selected candidates - by directly comparing their energies.

The 2- and 4-partite constraint used in EM was further assessed by
standard MC annealing simulations. We checked all interesting regions
of the phase diagram by running several MC simulations starting from
different initial configurations (both minimal or close-to-minimal EM
structures and random configurations) and used the annealing method
with exponentially decaying temperature profile. We tried several
system sizes with $2\times 2$, $4\times 4$ and $16\times 16$ number of
traps using periodic boundary conditions. As a result, we identified
and confirmed all shapes obtained in EM. However, due to the tiny
energy differences and large energy barriers between local minima,
long relaxation times (i.e., 10$^8$ MC steps for $4\times 4$ system
array) were needed to obtain exactly the same orientational orders as
in EM. Hence, the system easily gets trapped at low annealing
temperature in some local metastable phase composed of domains of
various ground state patterns \cite{rque10}.

\section{Square and rectangular lattice confinement}
\label{sec:square}

\begin{figure}[ht]
\includegraphics[width=0.45\textwidth]{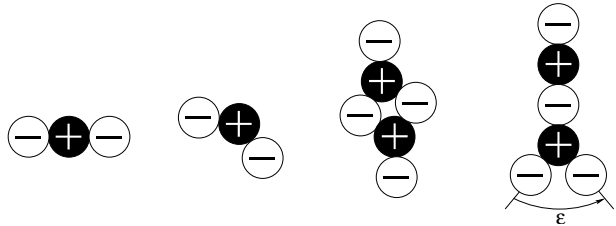}
\caption{Summary of the most frequent composite clusters encountered
  in our study (from left to right): straight trimer, banana ${\cal
    B}$, grape $\cal G$, rocket $\cal R$.}
\label{fig:summary}
\end{figure}

\begin{figure}[ht]
\includegraphics[width=0.65\textwidth]{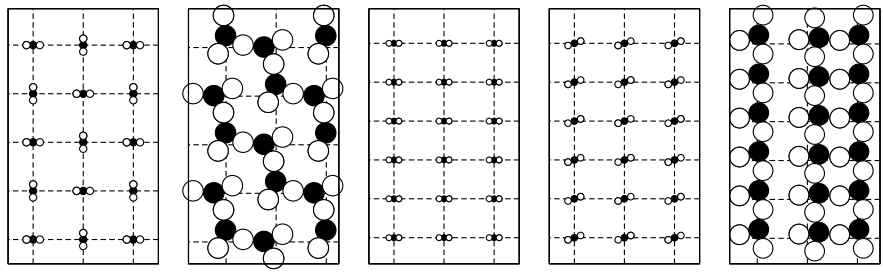}
\caption{Sketch of the different ordering patterns that will be
reported (from left to right): anti-ferro ${\cal A}$, chain $\cal C$,
ferro $\cal F$, tilted ferro $\cal F^*$ and ladder $\cal L$. Dashed
lines represent a square (or rectangular) lattice of traps. }
\label{fig:summary_phases}
\end{figure}

Before embarking on a detailed study, we first give an overview
(Fig. \ref{fig:summary}) of the various composite objects that we have
observed in our simulations. When the traps were well separated, we
observed isolated trimers, either straight or banana-shaped ${\cal
  B}$. For very small trap separation $l$, percolated chain-like
structures $\cal C$ rather than clusters are stable. In
the intermediate region ``super-clusters'' composed of six colloids
appear. They are formed by two neighbouring trimers forming the
super-cluster by minimizing their total energy (paying the penalty due
to the confinement while reducing the electrostatic interaction
energy) \cite{foot_cluster1}. We observed two typical shapes and
coined them grape $\cal G$ and rocket $\cal R$. In
Fig. \ref{fig:summary_phases} we further provide a summary of
different orientational orderings and propose a nomenclature to
classify them.

\subsection{Square lattices}
\label{ssec:square}

\begin{figure}[ht]
\includegraphics[width=0.75\textwidth]{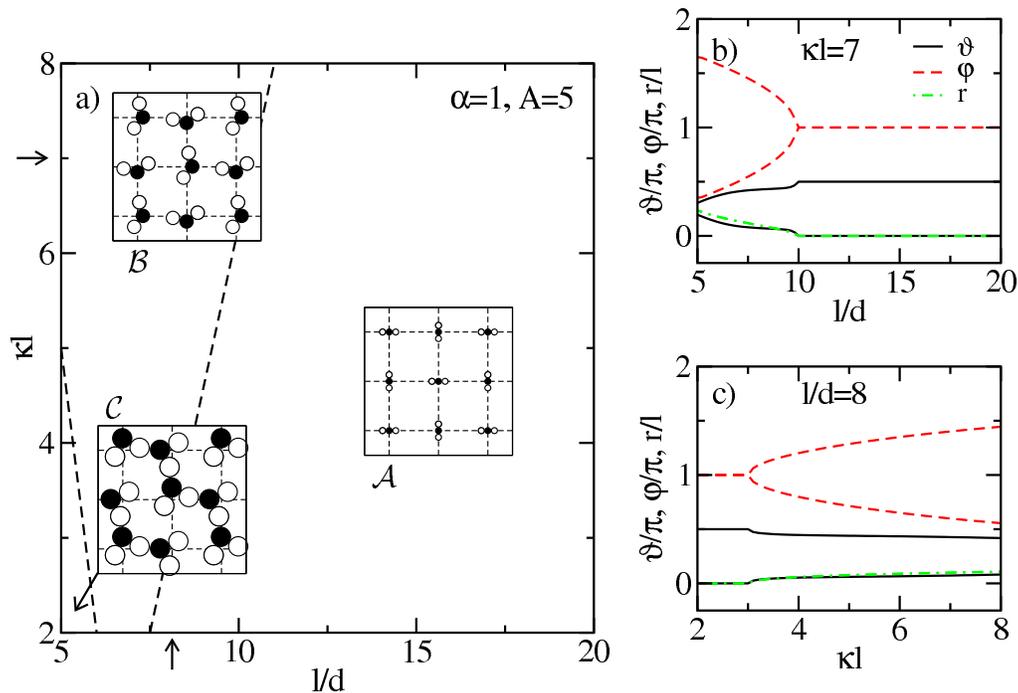}
\caption{(Color online) {\bf a)} Ground state phase diagram at a
  relatively strong substrate potential $(A=5)$ for colloidal trimers
  on a square lattice. Typical snapshot configurations are calculated
  using $(l/d,\kappa l)=(7,2)$ for the ${\cal B}$ phase, $(15,4)$ for
  the $\cal A$ structure, and $(5,2)$ for the $\cal C$ phase. The
  lines between the different phases are shown as a guide to the
  eye. {\bf b)} Values of the three quantities characterizing the
  cluster shape: (red dashed (upper): $r$, black solid: $\vartheta$
  and green dot-dashed (lower): $\varphi$), as a function of the ratio
  $l/d$ at constant $\kappa l=7$. {\bf c)} Values of the same three
  quantities as a function of $\kappa l$ at constant $l/d=8$.}
\label{R1}
\end{figure}
 
\begin{figure}[ht]
\includegraphics[width=0.75\textwidth]{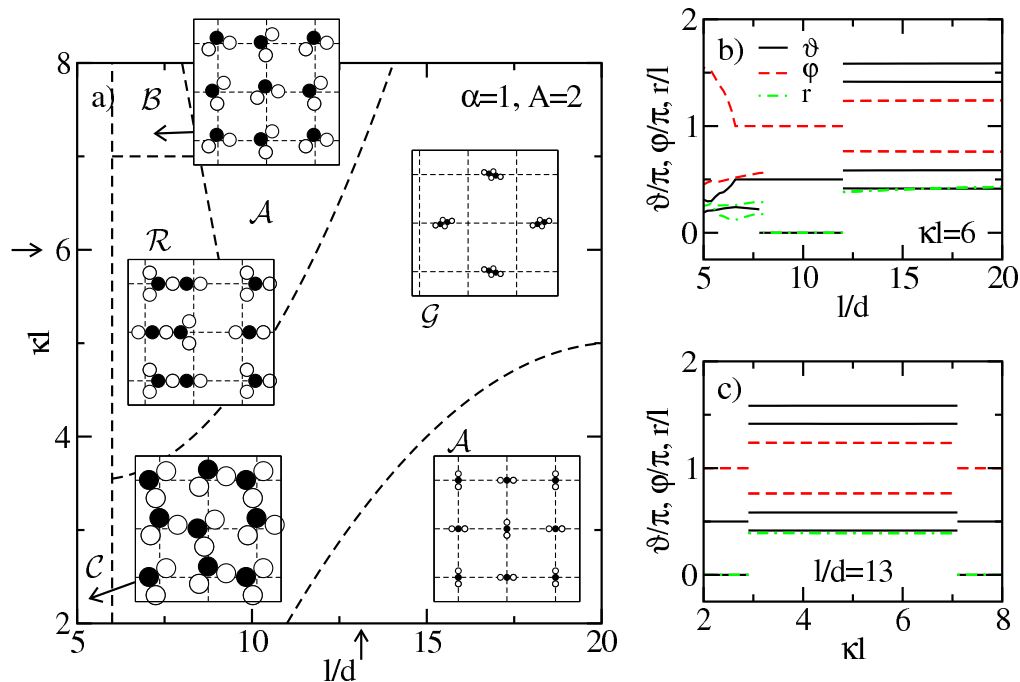}
\caption{(Color online) Same as Fig. \ref{R1} but for a relatively
  weak substrate potential $(A=2)$. {\bf a)} Typical snapshot
  configurations are calculated using $(l/d,\kappa l)=(7,8)$ for $\cal
  B$, $(7,4)$ for $\cal R$, $(5,2)$ for $\cal C$, $(20,8)$ for ${\cal
    G}$, and $(15,2)$ for $\cal A$ phase.}
\label{R2}
\end{figure}

Figures \ref{R1} and \ref{R2} show the type of orderings which occur
on a square lattice ($\alpha=1$), at two different pinning amplitudes
(embodied in the parameter $A$). On panel a) we show the phase diagram
separating the regions of stability of phases and the corresponding
snapshots of the typical cluster arrangements. The panels b) and c)
display the variation of three quantities characterizing the cluster
shape and their orientational ordering. The transitions from straight
to bent shapes are clearly seen on bifurcating red curves. It should
be stressed that $\phi$ and $2\pi-\phi$ represent the same shape and
the two branches do not indicate two different structures. The black
curves showing the cluster orientation have two branches with
$\vartheta_1$ and $\vartheta_2$. The special case is the ${\cal A}$
(antiferro) phase, where $(\vartheta_1,\vartheta_2)=(0,\pi/2)$. For
straight clusters the angles $\vartheta$ and $2\pi-\vartheta$ are
equivalent, but this is no longer true if they are bent. Therefore,
the region with four distinct branches on Fig. \ref{R2} represents a
4-partite structure with $\vartheta_1 ... \vartheta_4$. The values of
$r$ (green curves) tell us whether the clusters are displaced from the
trap minima ($r\ne 0$) or not ($r=0$). The values of the fourth
defining quantity $\gamma$ proved rather uninformative and were not
included in the figures for clarity reasons.

At the stronger confinement (Fig. \ref{R1}) we found three phases:
isolated triplets in the anti-ferro like arrangement $\cal{A}$, bent
(banana) triplets $\cal B$ whose angular arrangement varies
continuously to anti-ferro $\cal{A}$ upon increasing $l/d$, and a
percolated chain pattern ($\cal C$) at very low $l/d$ and small
screening (lower left corner of Fig. \ref{R1}a)). At weaker
confinement (Fig. \ref{R2}), two additional phases appear: rocket
$\cal R$ and grape $\cal G$, which are both non-percolating, and made
up of the repetition of 2-triplet (6 colloids) objects. 

In Fig. \ref{R1}, all ground state configurations have a 2-partite
lattice structure, which has also been confirmed by the unconstrained
MC annealing simulations. In the limit of small macro-ions, $l/d\gg
1$, we find stable isolated triplet configurations with quasi
degenerate orientational preferences. For example, the anti-ferro
$\cal A$ phase is slightly more favorable compared to the tilted ferro
${\cal F}$* phase: at the $(l/d,\kappa l)=(15,4)$ point in the phase
diagram the corresponding energies are $E_{\cal A}/K=-2.4250$ and
$E_{\cal F^*}/K=-2.4247$. 

 In Fig. \ref{R2} there are regions in the phase diagram where the
4-partite configurations appear to be favored over bipartite. The
energy difference between 2- and 4-partite ordering, however, is
extremely small. For example, in the representative rocket $\cal R$
and grape $\cal G$ phases shown in Fig. \ref{R2}(a) the energies are
respectively $E_{4p}/K=-0.5888$, $E_{2p}/K=-0.5870$ and
$E_{4p}/K=-2.723145$, $E_{2p}/K=-2.723143$.  Such small differences,
$E_{4p}-E_{2p}$, imply that these ground state configurations become
unstable at small but finite temperatures and the formation of
domains
 is expected in experiments and simulations. Similar effect
is expected
 also for the $\cal A$ ground state, which should
interfere with
 nearest (in energy) $\cal F$* phase, as in the
previous $A=5$ case in
 Fig. \ref{R2}. On the other hand, the grape
$\cal G$ phase in the
 $l/d\gg 1$ limit is well separated from the
competing $\cal A$ phase
 ($E_{\cal A}/K=-2.645310$ for the same
$\kappa l$ and $\kappa d$
 values).
 
\subsubsection{Antiferro-like phase}

\begin{figure}[ht]
\includegraphics[width=0.4\textwidth]{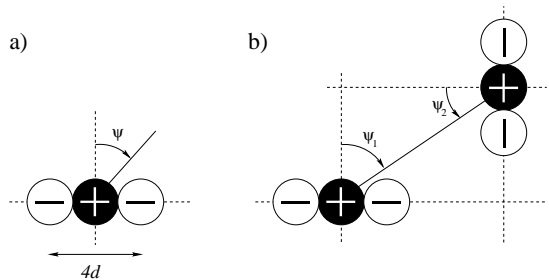}
\caption{Definition of the notations used to compute triplet-triplet
interaction in the (aligned) straight case.}
\label{fig:triplet}
\end{figure}

The occurrence of the $\cal A$ phase on the right hand-side of
Figs. \ref{R1} and \ref{R2} can be rationalized following similar
lines as in Ref. \cite{Agra}. For large enough values of $\kappa l$
and even larger $l/d$, it is legitimate to restrict the summation in
Eq. (\ref{eq:energy}) to nearest neighbor traps (see \cite{ElShawish}
for a more complete discussion). It is also legitimate to assume that
triplets align in straight configurations. The question is therefore
to understand the relative orientation of triplets, from trap to
trap. Elaborating on the remark, put forward in \cite{JPCM1,EPJE},
that a non isotropic charge distribution in an electrolyte, creates an
electric potential that is anisotropic at all scales and that does not
have any simple `multipolar' symmetry, we write the angular dependence
of the far field potential of a triplet as \cite{Agra}
\be
  V_{tr}=2 \cosh(2 \kappa d \sin\psi) -1
\ee
where the direction $\psi$ is defined in Fig. \ref{fig:triplet}-a).
Consequently, the relevant angular dependent triplet-triplet
interaction potential may be written (see Fig. \ref{fig:triplet}-b)):
\be
  V_{tr/tr}=\left[2 \cosh(2 \kappa d \sin\psi_1)-1\right]
  \left[2 \cosh(2 \kappa d \sin\psi_2) -1\right]. \label{eq:triplet-triplet}
\ee
On a square lattice, and assuming bipartite structure (which is backed
up by simulations), we then have to consider triplet-triplet
interactions along both principal axis of the lattice. This leads to
the following energy
\begin{eqnarray}
&  \left[2 \cosh(2 \kappa d \sin\psi_1) -1\right] \left[2 \cosh(2
  \kappa d \sin\psi_2) -1\right] \nonumber \\
  + &\left[2 \cosh(2 \kappa d
  \cos\psi_1) -1\right] \left[2 \cosh(2 \kappa d \cos\psi_2)
  -1\right].
\end{eqnarray}
The minimum is reached for $(\psi_1,\psi_2)=(0,\pi/2)$, irrespective
of $\kappa d$, which exactly coincides with the $\cal A$ phase
configuration. The situation changes on a rectangular lattice, see
below.  On the other hand, the $\cal A \to B$ transition line observed
in Fig. \ref{R1} can be explained as a one-cluster effect: in the
absence of confinement, the trimers take on the straight shape due to
the repulsion between the two positively charged colloids. However,
by increasing the confinement or because of enhanced screening, the
relative importance of the electrostatic repulsion decreases. We
expect that for $A$ and $d$ fixed, there exists a critical inverse
screening length $\kappa^*$ beyond which the straight trimer bends
(and ultimately forms an equilateral triangle if $\kappa$ is very
large). Since it is a one-triplet instability, $\kappa^*$ should be
independent of lattice spacing $l$ which translates into a straight
line with slope $\kappa^* d$ in a $(\kappa l,l/d)$ plot such as
Fig. \ref{R1}(a). This corroborates our numerical finding with a
straight separatrix between $\cal A$ and $\cal B$ phases.  For $A=2$,
the argument does not hold since no transition from straight to bent
shapes takes place.

\subsubsection{Grape phase}
The stability of the grape phase $\cal{G}$ can be rationalized by
comparing the self-energy per colloid 
\be
 E_{self}=\frac{d}{6} \sum_{j\ne i=1}^6 s_{ij}\frac{e^{-\kappa r_{ij}}}{r_{ij}}
\ee
of different composite complexes in the absence of any trap potential.
\begin{figure}[ht]
\includegraphics[width=0.5\textwidth]{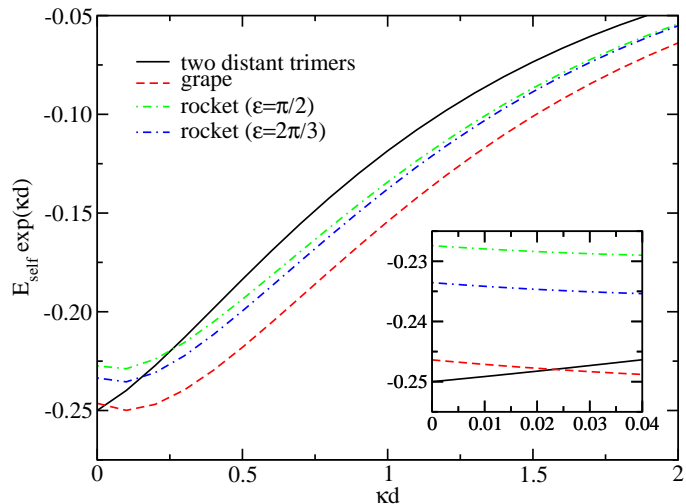}
\caption{(Color online) Coulombic self-energy of 6 colloids (2
  positively, and 4
 negatively charged) forming various
  complexes. }
\label{fig:self}
\end{figure}
In Fig. \ref{fig:self} we have compared the self-energies of the
following six-colloid objects: grape, rocket (with various angles
$\epsilon$, as defined on Fig. \ref{fig:summary}) and two distant
trimers. Among these, the two distant trimers are the most stable in
the bare Coulomb limit ($\kappa\to 0$) with $E_{self} \to
-1/4$. However, as soon as $\kappa d > 2.2\,\cdot 10^{-2}$, a fairly
small number, the grape geometry is found to be the optimal
one. Therefore, in almost the entire range of relevant values, the
grape is energetically the most stable isolated object consisting of
six colloids. Of course, on a corrugated substrate, the confining
potential influences the above scenario by penalizing extended
structures, and the inter-cluster interaction affect the picture as
well, however, there is still a substantial region on Fig. \ref{R2}
and \ref{R3} (see below) where the grape phase remains stable.

\subsection{Rectangular lattice of traps}
\label{ssec:rectangular}

\begin{figure}[ht]
\includegraphics[width=0.75\textwidth]{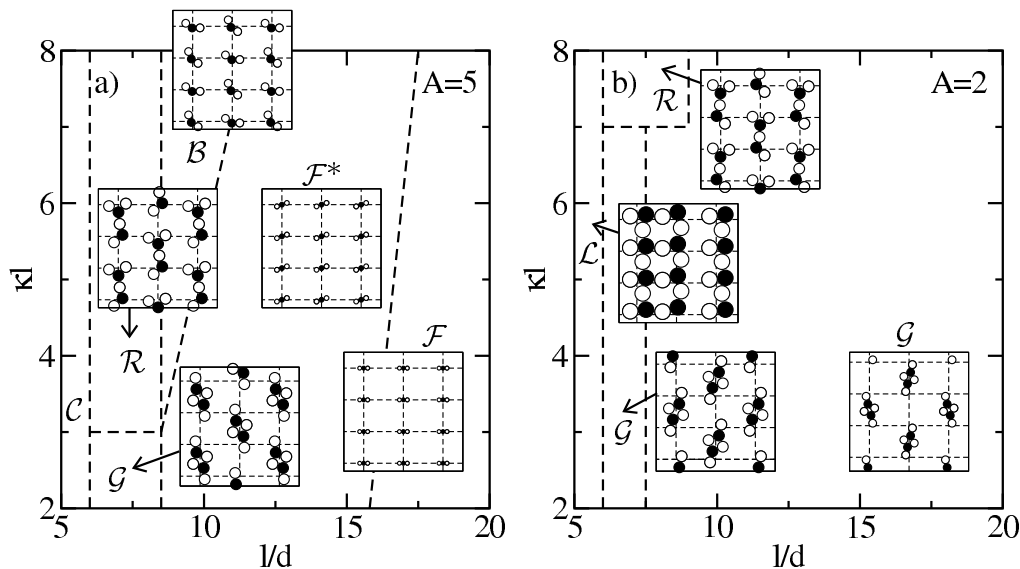}
\caption{ Ground state phase diagram for colloidal trimers on a
  rectangular
 lattice with $\alpha=0.8$ and {\bf a)} $A=5$ or {\bf
    b)}
 $A=2$. Typical snapshot configurations in {\bf a)} are
  calculated
 using $(l/d,\kappa l)=(10,8)$ for $\cal B$, $(7,4)$ for
  $\cal R$,
 $(7,2)$ for $\cal G$, $(15,6)$ for $\cal F$*, and
  $(20,2)$ for $\cal
 F$ phase; in {\bf b)} $(l/d,\kappa l)=(7,8)$
  for $\cal R$, $(5,6)$ for
 $\cal L$, $(7,2)$ for 2-partite ${\cal
    G}$, and $(10,2)$ for 4-partite
 ${\cal G}$ phase.}
\label{R3}
\end{figure}

Distorting lattice geometry by changing the aspect ratio away from the
square shape ($\alpha\equiv l_y/l_x <1$) induces new orderings to the
list of ground state configurations reported above. Figure \ref{R3}
shows ground state phase diagrams for $\alpha=0.8$ and two pinning
potentials, where three new phases emerge with respect to the square
($\alpha=1$) case: a ferro $\cal F$ phase and a tilted ferro $\cal F$*
phase, which evolves into a familiar banana $\cal B$ phase at smaller
$l/d$ and $A=5$, and a percolating ladder $\cal L$ phase in a weaker
confinement ($A=2$). The latter one 
consists of complex subunits where each trimer touches with three (out
of four) neighboring trimers. Both pinning regimes involve also a
distorted (or bent) rocket $\cal R$ phases, which we identify as a
variant of the existing ($\alpha=1$) phase rather than a new phase
itself, and a grape $\cal G$ phase with 2- and 4-partite
structures. Actually, in Fig. \ref{R3}(b), almost the whole parameter
space is minimized by a 4-partite grape $\cal G$ configuration
\cite{2-4}.

On a rectangular lattice, we can follow a similar line of reasoning as
above in section \ref{ssec:square}, where the occurrence of the
${\cal A}$ phase was justified. In contrast to the situation on the square
lattice, on the rectangular lattice the number of nearest neighbor
traps decreases to two and in order to assess the stability of the
${\cal F}$ phase, it is sufficient to consider a single
triplet-triplet interaction of the form
Eq. (\ref{eq:triplet-triplet}). In this case, the lowest energy
configuration corresponds to parallel triplets ($\psi_1=\psi_2=0$)
forming a ferromagnetic-like arrangement along the longest direction
of the lattice unit cell. This is indeed what is observed in
Fig. \ref{R3}(a) (see the $\cal F$ region on the right hand side), but
not in Fig. \ref{R3}(b), because in this latter case, confinement
strength is too weak to prevent neighboring traps from sharing their
macroions. We further note that the $\cal F^* \to B$ transition
visible in Fig. \ref{R3}(a) falls in the same category as the $\cal A
\to B$ seen in Fig. \ref{R1}: it is a one-cluster instability due to
the competition between Coulomb and confinement forces. It is hence
not surprising to find that the slopes of both separatrices are close:
they are quantified by the same critical screening coefficient
$\kappa^*$ introduced above.

\section{Towards a more realistic trap potential}
\label{sec:realtrap}

On a realistic trap, when allowing the particles to hop, we can expect
to significantly extend the stability of the grapes, at least as soon
as $\kappa d$ exceeds a small threshold. This is supported by the
evaluation of the self energies on Fig. \ref{fig:self} and we further
analyzed the stability of the grape phase by explicitly introducing a
cosine potential, which provides a reasonable approximation for a
realistic optical confinement \cite{RO}:
\begin{equation}
   V_T^{real}=-\frac{A_r}{2 \pi^2} \, (\cos(2\pi x/l)+\cos(2\pi y/\alpha l))\;.
\label{Vreal}
\end{equation}
The prefactor $2 \pi^2$ is chosen such that both realistic and simplified
potentials behave in the same way, 
as $(\delta/l)^2 = (x^2+y^2)/l^2$ close to the trap minimum,
when one chooses $A=A_r$.

\begin{figure}[ht]
\includegraphics[width=0.65 \textwidth]{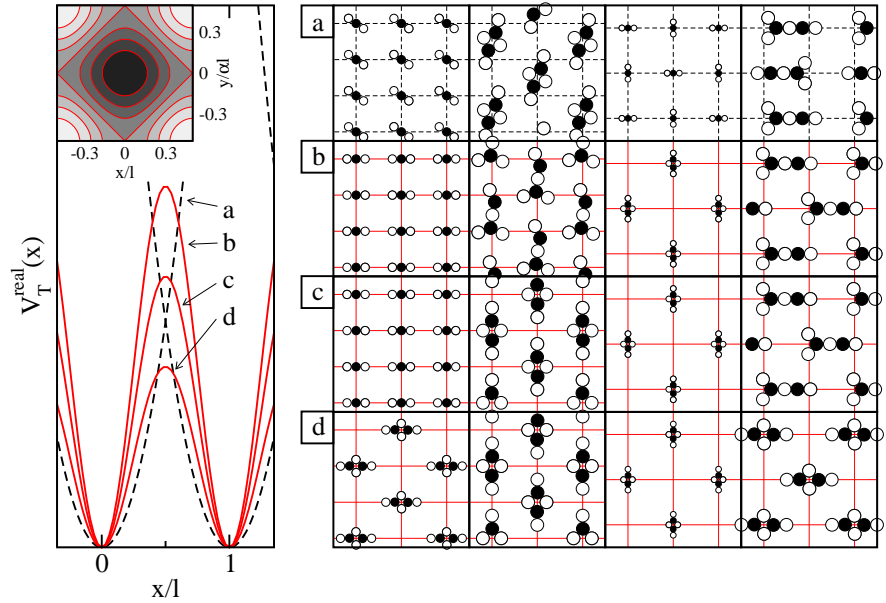}
\caption{(Color online) Comparison of structures calculated within an
  isotropic
 parabolic potential (a) and realistic cosine potential
  $V_T^{real}$
 (b-d) with three neighboring magnitudes $A_r$ chosen
  to roughly
 correspond to the parabolic potential at the trap
  midpoint $x/l=0.5$
 (left panel). In (a) the parameter values are,
  from left to right: 
 $(l/d,\kappa
 l,A,\alpha)=(10,4,5,0.8)$
  ($\cal F$* phase), $(7,2,2,0.8)$ ($\cal G$
 phase), $(15,4,5,1)$
  ($\cal A$ phase), and $(7,4,2,1)$ ($\cal R$
 phase). In (b-d) we
  use the same parameters as in (a) except
 $V_T^{real}$ is used
  instead of parabolic trapping with
 $A_r=20, 15, 10$ for $A=5$ and
  $A_r=8, 6,
 4$ for $A=2$.}
\label{R5}
\end{figure}

In Figs. \ref{R5}(b,c,d) we show ground state configurations for
various pinning strengths $A_r$ using the same parameter values $(l/d,
\kappa l,\alpha)$ as in Fig. \ref{R5}(a) where we show the four most
representative parabolic-trap configurations. We choose $A_r$ such
that both potentials become similar in amplitude at the point of
maximum $V_T^{real}$ (see left panel of Fig. \ref{R5}). As $A_r$ gets
smaller (going from Fig. \ref{R5}(b) to Fig. \ref{R5}(d)), the optimal
configurations indeed evolve to the grape $\cal G$ geometry. In the
limit $A_r\to A$, i.e., when the potentials share the same behaviour
close to the trap center, the lowest configurations are always grape
$\cal G$ phases. Although having the same internal structure, these
grapes are formed in realistic trap minima rather than at trap
midpoints. In addition, the orientation of grapes is selected by the
symmetry of the underlying realistic potential $V_T^{real}$, which is
lowered with respect to the isotropic parabolic trap (the inset in the
left panel of Fig. \ref{R5} shows the equipotential lines of
$V_T^{real}$). In fact, the realistic trap square symmetry is manifested in
all configurations shown in Figs. \ref{R5}(b,c,d).

From Fig. \ref{R5}, we see that switching to realistic trap potential
($V_T^{real}$) preserves some typical cluster shapes (e.g., straight
trimer, grape $\cal G$, rocket $\cal R$) from the parabolic case,
while the positional or/and orientational order is generally
altered. In a simple picture, the full consistency of both potentials
(if any) would be found near the bottom of the trap where the
realistic potential becomes isotropic, i.e., for $l/d\gg 1$. Since
small $A_r$ favor grape configurations that violate both the
positional and orientational orders of the parabolic case, the
consistency should be rather searched in the opposite limit of strong
$A_r\gg 1$, where each trap should host exactly 3 colloids. In
particular, the most promising phases are $\cal F$ and $\cal A$ phases
since they also obey the symmetry constraints of $V_T^{real}$. To
avoid bending of extended trimers in these two phases, we also
anticipate small screening, i.e., $\kappa d\ll 1$, otherwise a banana
$\cal B$ shape might become more favorable.

\begin{figure}[ht]
\includegraphics[width=0.55\textwidth]{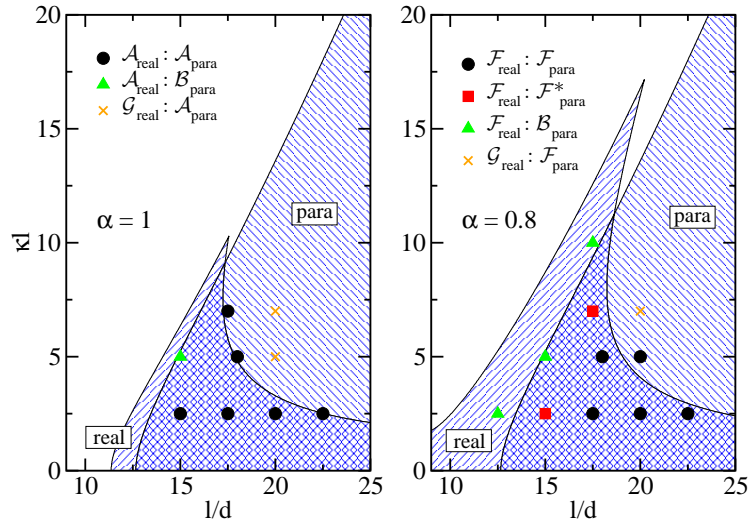}
\caption{(Color online) Single-trap calculations for parabolic (para)
  and realistic
 (real) trap potentials using $A=A_r=50$ and
  $\alpha=1$ (left panel)
 and $\alpha=0.8$ (right panel). Shaded
  areas denote regions in
 parameter space where straight-trimer has
  lower energy then a banana
 $\cal B$ and grape $\cal G$
  clusters. The symbols are comparing full
 energy-minimization
  calculations for both trap potentials
 considering all neighboring
  traps: black disks denote exact
 consistency while the other
  symbols indicate that different
 phases are found with the
  `realistic' and parabolic confinement 
 potentials.}
\label{consist}
\end{figure}

Following these criteria and using a single-trap approximation (which
is justified for small clusters), we identified regions in the
$(\kappa l, l/d)$ parameter space (see shaded areas in
Fig. \ref{consist}) where one straight-trimer is energetically
favorable against the banana $\cal B$ and grape $\cal G$
structures. In Fig. \ref{consist} we show such regions for realistic
and parabolic potentials for a square and rectangular case (note that
$\alpha$ has no effect in the parabolic approximation). The overlap
between them gives an estimate where the consistency for $\cal F$ (in
$\alpha=0.8$) and $\cal A$ (in $\alpha=1$) should be searched within a
proper calculation including all neighboring traps. We have checked
this at few points of the phase diagram (symbols in
Fig. \ref{consist}) and indeed found consistency: denoted by black
circles in Fig. \ref{consist} are the results where the all-neighbour
calculation predicts the same phase ($\cal F$ in Fig. \ref{consist}a
and $\cal A$ in Fig. \ref{consist}b) for both potentials. These results
agree with the single trap prediction. To the left from the
overlapping region (at smaller $l/d$), the inclusion of the
neighboring traps introduces bending of the trimers and thus prefers
banana $\cal B$ phase over the $\cal F$ or $\cal A$ (see green
triangles in Fig. \ref{consist}). This effect is more visible for
parabolic potential since it is stronger than realistic when
$A=A_r$. The above calculations were performed for a rather
strong pinning strength ($A=A_r=50$); however, we have checked and
found $\cal F$ and $\cal A$ consistency also for weaker trap
potentials, with $A=A_r\gtrsim 10$.

\section{Applying an external electric field}
\label{sec:electric}

Since most of the observed phases are generally  close in energy,
it is tempting to explore how an external perturbation could change
the phase behaviour. We studied the response of the system to a static
uniform electric field ${\bf E}$ with the in-plane direction. Our
primary purpose was to numerically investigate the possibility of
external pattern switching between the known phases, which might be
 appealing for experimental realizations. The ability to
create and control colloidal crystal structures indeed has a wide range of
applications in photonic and phononic materials, optical switches,
photonic band-gap materials, and self-assembly of nanostructures
\cite{Bunch1}.

In addition to either parabolic or cosine confinement, we use the
following field potential at position $(x,y)$:
\begin{equation}
  V_E=\pm E_0(x\cos\beta + y\sin\beta)/l,\label{eq:field}
\end{equation}
where $E_0$ is a dimensionless field strength measuring the relative
importance of external drag forces over Coulombic forces, and $\beta$
is an angle between ${\bf E}$ and the longest lattice principal
axis. The "$\pm$" sign selects between positively and negatively
charged colloids, respectively. The amplitude $E_0$ is considered
small enough so that system acquires equilibrated ground state
configuration. This requirement is easily fulfilled in a parabolic
trap approximation where, in principle, we never reach stationary
regimes with drifting colloids \cite{Olson1} even for largest values
of $E_0$.
However, the opposite may happen in a realistic trap confinement
\cite{Olson1} where at critical $E_0$ either cluster hopping or its
breaking into smaller constituents takes place. We note that in the
later case the critical field can be expressed analytically by
$E_{c,2}=(1+2\kappa d)\exp(-2\kappa d)/(2d/l)^2$ above which the two
oppositely charged colloids (a dimer) unbind. The breaking of a trimer
occurs at slightly lower fields that can be estimated numerically by
$E_{c,3}\sim 0.8 E_{c,2}$.

\begin{figure}[ht]
\includegraphics[width=0.7\textwidth]{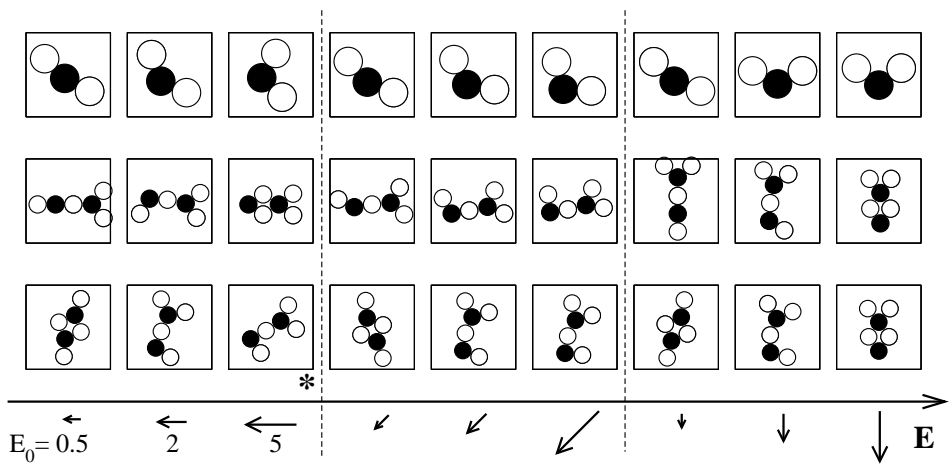}
\caption{Application of the external electric field ${\bf E}$ on
  colloidal clusters shapes in parabolic pinning regime for three
  field strengths $E_0$ and directions depicted at the bottom of the
  figure. The other parameters are: $(l/d,\kappa
  l,A,\alpha)=(10,4,5,0.8)$ in first row, $(7,4,2,1)$ in second row,
  and $(7,2,2,0.8)$ in third row. The asterisk symbol indicates that a
  cluster (when viewed periodically on a lattice) forms a percolated
  structure along the field direction.}
\label{R7}
\end{figure}

We examined the influence of ${\bf E}$ on the selected phases shown in
Secs. \ref{sec:square} and \ref{sec:realtrap} using the EM and the gradient
descent method. In Fig. \ref{R7} we show the results (one-trap snapshots) for
parabolic
trapping. Apart from rather trivial effects, such as bending
(formation of the $\cal B$ shape out of the straight trimer, or a
rocket $\cal R$ deformation for $E_0\le 2$), we identified also more
interesting effects at strongest $E_0=5$: a new shape evolved from
$\cal R$ as well as $\cal G$ structure, and an elongation of the
modified $\cal G$ complex along the field direction lead to the
formation of the percolated chain structure (denoted by asterisk
symbol in Fig. \ref{R7}).

\begin{figure}[ht]
\includegraphics[width=0.6\textwidth]{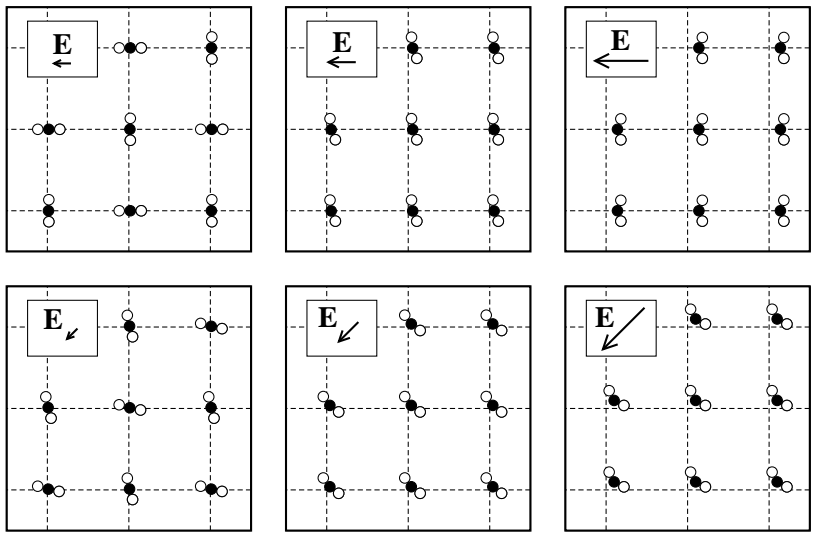}
\caption{Influence of an external electric field ${\bf E}$ on the
  $\cal A$ phase within the parabolic trap approximation for $(l/d,\kappa
  l,A, \alpha)=(15,4,5,1)$. The field directions and magnitudes are
  shown with arrows in the inset of each plot; arrow lengths
  correspond to magnitudes $E_0=0.5, 2, 5$ of
  Eq. (\ref{eq:field}). For $E_0=2$ and $\beta=0,\pi/4$ (the middle
  column) we observe a pattern switching to the $\cal F$* phase
  (actually, its close approximation, since the trimers here are
  slightly bent with $\varphi_{1,2}\approx 0.91\pi$).}
\label{R6}
\end{figure}
In a similar manner, we demonstrate in Fig. \ref{R6} the influence of
${\bf E}$ on a particular $\cal A$ phase, again within a parabolic
trap confinement; for $E_0=2$ and an electric field angle 
$\beta=0$ or $\pi/4$, we identify a
pattern switching to a (quasi) $\cal F$* phase. Here, the transition $\cal
A\to \cal F$* appears to be on orientational basis, with more or less minor
shape deformations.
Since the trimers of field induced $\cal F$*
phase are slightly shifted away from the trap center, the expected
bending due to electric field is (almost) compensated by the restoring
trap forces and the clusters appear almost straight.\\


We have  also explored the effect of  the external field within  the real trap
confinement. As found by Reichhardt et al in Ref.  \cite{Olson1}, the colloids
start to flow as soon as the electric field force becomes larger than the real
trap  confinement, but  here we  rather  focused on  static configurations  at
smaller  field strengths.  In fact,  in a  large enough  system,  any constant
electric field would  inevitably lead to ground states  with uneven population
of the traps,  i.e. the trimers would  prefer to move in the  direction of the
field to reduce the  total energy of the system, leaving some  of the traps at
the opposite end empty. However, even though such states might have the lowest
energy, they might  not be accessible in the  experiments and simulations. If,
for instance, the system is first let to equilibrate at ${\bf E}=0$ and then a
small enough electric  field is turned on, a local  rather than global minimum
is reached with  colloids unable to surmount the large  energy barriers on the
way to the global minimum. This is the regime we are interested in here.

\begin{figure}[ht]
\includegraphics[width=0.45\textwidth]{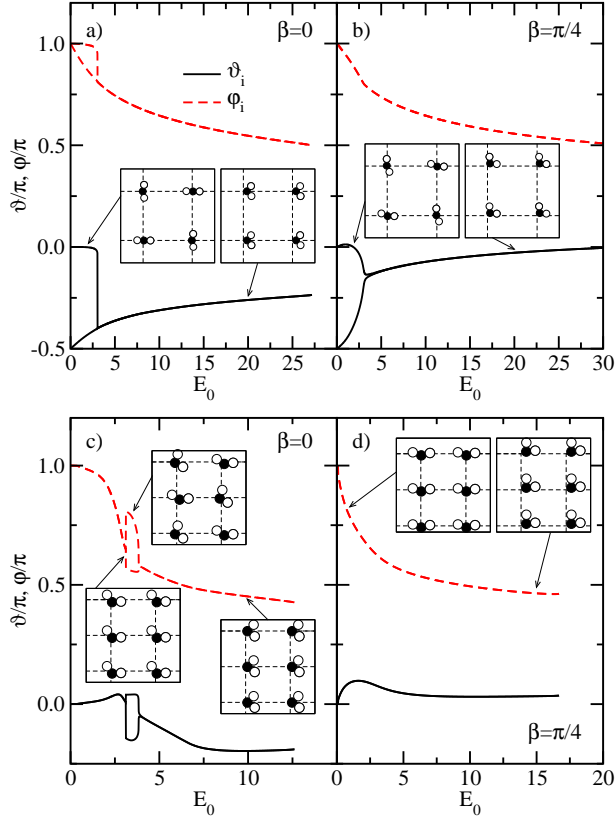}
\caption{(Color online) Local energy minima configurations represented
  by trimer angles $\vartheta_i$ (solid line) and $\varphi_i$ (dashed)
  as a function of external electric field $E_0$ using the cosine
  confinement potential Eq. (\ref{Vreal}). The index $i=1...4$
  represents the four different values for a 4-partite lattice
  assumption. The upper row is calculated from initial ($E_0=0$) $\cal
  A$ phase with
 $(l/d,\kappa l,A_r/(2\pi^2), \alpha) = (15,4,1.5,1)$
  for two field
 directions a) $\beta=0$ and b) $\beta=\pi/4$. In the
  lower row the
 initial ($E_0=0$) state is the $\cal F$ phase with
  $(l/d,\kappa
 l,A_r/(2\pi^2), \alpha)=(10,4,0.75,0.8)$ and field
  directions are
 again c) $\beta=0$ and d) $\beta=\pi/4$. In all the
  plots we used
 field resolution $\delta E_0=0.01$. A few
  representative snapshots are
 added for illustration (see the
  supporting material \cite{EPAPS} for the animated cartoons).}
\label{R8}
\end{figure}

To search for nearest local energy minimum with an increasing
electric field, we introduced a gradient descent method (i.e., a
method of steepest descent) on a 4-partite lattice (16 model
parameters, see Fig. \ref{S}) which brings a system continuously from
an initial zero-field ground state to a finite-field metastable state
using a finite step $\delta E_0$ (at fixed field direction
$\beta$). Depending on the amplitude of the increment
$\delta E_0$, the switching of
the field may be interpreted as fast or slow, something that can also
be tuned in an experimental setup and may affect the outcome. In our
calculations we tried three step sizes, $\delta E_0=0.01, 0.1, 1$, and
found basically no differences in the resulted configurations.

In Fig. \ref{R8} we show the evolution of $\cal A$ and $\cal F$ phases
with increasing $E_0$ along two field directions using highest field
resolution $\delta E_0=0.01$. Focusing first on the $\cal A$ phase
(upper row in Fig. \ref{R8}), we identify at $E_0\approx 3$ (for both
field directions) a rapid phase transition from a 2- to 1-partite
phase, which can be also viewed as a pattern switching to an
approximate $\cal F$* phase (with the internal angle being
$\varphi\approx 0.8\pi$ rather than $\pi$), similar to what we found
in the parabolic confinement (middle column in Fig. \ref{R6}). This
tendency of reducing the ``partiteness'' of the phase by increasing the
external uniform field seems natural. However, the opposite situation
is observed in the evolution of the $\cal F$ phase, shown in
Fig. \ref{R8}(c), where in a short interval, $3\lesssim E_0\lesssim
4$, the metastable phase adopts a higher, 2-partite $\cal B$ structure. 
This rather surprising effect reflects the richness
of the system studied, where the observed phases differ only slightly
in energy and can therefore interchange easily under
external perturbation. At larger field strengths again a uniform
pattern takes place.

We note that a similar field-evolution study in cosine trap
confinement was made also for two representative $\cal G$ and $\cal R$
phases with 6 colloids per trap, but we found no striking effects
apart from trivial bending of the clusters, which was already observed
in a parabolic confinement (see, e.g., Fig. \ref{R7}). 
These results are included in a form of animated cartoons available
as online supporting material \cite{EPAPS}.

\section{Conclusion}
\label{sec:concl}

 In summary, we have studied ordering of colloidal ionic trimers on
square lattices. We observed a rich variety of crystal structures
including crystals made of complex 6-colloid subunits. We have
discussed their stability and the possibility for pattern switching
by
 external control, which might be appealing for experimental
realisations and applications.
 
 In case of the isotropic
confinement potential, the symmetry is broken
 by the Yukawa
interactions only. The orientational 
 ordering is of `intrinsic
nature' and does not stem from an 
 anisotropy present in the
confinement potential itself. 
 Three colloids can form a straight
line or a bent banana
 shaped trimer. We have identified the
parameter range where both are
 stable and the types of their
orientational ordering. In the low
 $\kappa$ and small $l/d$ part of
the phase diagrams, where the
 inter-cluster interactions typically
outweigh the confinement, the
 neighboring trimers cluster into
larger complexes consisting of six
 colloids; their typical shapes
being grape and rocket. At even smaller
 $l/d$ and $\kappa$, the
colloids form percolated structures. Here,
 however, many-body
effects \cite{3-body_Brunner} are expected to become significant and
our assumption of the pairwise additivity of the colloid-colloid
interaction becomes questionable
\cite{JCP_manyB,Reinke_PRL_Cauchy}. We have shown that most of the
phases observed in the isotropic confinement persist also in the more
realistic anisotropic external potential, provided that the
confinement strength is large enough. However, their orientational
ordering in the latter case is different due to the additional
symmetry breaking by the laser field. The rich selection of the
ground
 state structures and the relatively small energy differences
between
 them enable efficient control over the structures by
applying external
 fields. We have analysed the effect of a small
external electric field
 and have shown that it can be used as a
pattern switching tool.
 
 Not addressed here, but a subject for
future work, is the thermal
 behaviour of the system. 
 Another
question lies
 is the effect of the interaction potential details on
the stability of
 the observed structures, and in 
 the many-body
interactions among colloids that are expected to be
 important at
high densities. Finally, 
 in addition to systems with oppositely
charged
 colloids, externally
 driven superparamagnetic colloids
\cite{col.membranes} and 
 colloids interacting via Casimir
interaction in critical binary
 mixtures \cite{Bechinger_Casimir} may
offer instances where the
 present predictions --or adaptations
thereof-- could be realized.
 

{\bf Acknowledgments: } We acknowledge the support of the bilateral
program Proteus of the Slovenian Research Agency and the French
Minist\`ere des Affaires Etrang\`eres et Europ\'eennes, the support of
the Slovenian research agency through the grant P1-0055, and the
support of the EU through the ARG (ERC-COLSTRUCTION 227758) and
7$^{th}$ Framework Programme (ITN-COMPLOIDS 234810).\\



\begin{thebibliography}{99}
\bibitem{rque}
As will be discussed later, our model differs in some aspects from 
the one considered in Ref. \cite{Sarlah}, where only discrete
orientations were considered for the colloidal $n$-mers.
Within such a framework, some approximations, backed up by Monte Carlo
simulations, have been put forward
to unravel the effect of temperature \cite{Sarlah}.
\bibitem{rque10} We stress that our system exhibits at least two energy
  scales: the larger one is defined by the attractive Coloumb interaction
  between two touching colloids and is of the order $1/2d$, while the smaller
  one is the Yukawa interaction between the traps of the order $\exp(-\kappa
  l)/l$ and is responsible for orientational order. The $4p$-dimensional
  energy surface is therefore composed of deep minima each of which has again
  many shallow minima.
\bibitem{foot_cluster1} This mechanism somehow resembles the
  strain-induced colloidal clustering on periodic (triangular)
  substrates with more traps than colloids \cite{Bleil}. Another
  somewhat related phenomenon is the appearance of the so-called
  cluster crystals in systems with soft repulsive potentials
  \cite{Mladek,Likos_clusters,Osterman1,Julia_jpcm}, although the
  present system is more complex and a delicate interplay between the
  external confinement and the colloidal interactions governs its
  behaviour.
\end{thebibliography}
\end{document}